\documentclass{llncs}

\usepackage{url}
\usepackage{todonotes}
\usepackage{graphicx}

\title{Analyzing and Improving Eligibility Verifiability of the Proposed Belgian Remote Voting System}

\author{Jan Willemson\inst{1}\orcidID{0000-0002-6290-2099}}
\authorrunning{J. Willemson}
\institute{Cybernetica, Narva mnt 20, 51009 Tartu, Estonia\\
\email{jan.willemson@cyber.ee}}

\begin{document}

\maketitle

\begin{abstract}
    This paper discusses several recent electronic-paper remote voting hybrid schemes, concentrating more specifically on the proposal put forward for Belgian elections. We point to some problems in the proposal, and consider addition of blind signatures as one possible approach to dealing with the identified shortcomings. We discuss the concomitant updates from both the protocol and electoral processes point of view, arguing that our proposal would strike a better balance between different requirements. To the best of our knowledge, this is also the first proposal to use blind signatures in a paper-based voting system.
    \begin{keywords}
        Remote voting \and eligibility verification \and blind signatures
    \end{keywords}
\end{abstract}

\section{Introduction}

In our current increasingly mobile world, it becomes harder and harder to get all the eligible voters to physical polling stations for the act of voting on a single day~\cite{DBLP:journals/istr/Willemson18}. The recent COVID-19 outburst has only added to this problem~\cite{cotti2021relationship}. Hence, there is a definite need for reliable remote voting options. 

Two main approaches are available for this. The more established way is to send the ballots in via physical mail. For example in the 2020 U.S. presidential elections, 43\% of the voters cast their ballot by mail -- a number twice as high as four years earlier. Even though the COVID pandemic was definitely a major factor, the trend towards increasing voting  by mail has been observed for years in the U.S.~\cite{https://doi.org/10.48550/arxiv.2111.08662}

As an alternative, several countries like Switzerland~\cite{DBLP:conf/evoteid/HainesPT22}, Estonia~\cite{EHIN2022101718}, Norway~\cite{DBLP:conf/ev/StenerudB12}, Australia~\cite{iVote22}, France~\cite{DBLP:conf/evoteid/BlanchardGLSW22}, etc. have had elections with vote casting options over Internet.

Both of these approaches have their pros and cons. Internet voting can offer reliable vote transmission  and efficient tallying procedures. On the other hand, it has been criticized for implementation complexity, concentrating many risks into the central components, being hard to verify by an average citizen, etc.~\cite{halderman2016practical,simons2019internet,park2021going} 

Postal voting can be implemented without relying on any digital equipment on the client side, hence  being easy to understand, use and trust by the voter.
On the other hand, it is very hard to ensure authenticity and privacy of the voters, the postal channel is vulnerable to both integrity and confidentiality attacks, etc.~\cite{DBLP:conf/egov/KrimmerV05,DBLP:conf/spw/BenalohRT13}

Thus, it is natural to ask whether we could get the best of the both worlds without sacrificing too much in terms of residual risks. And indeed, starting from Benaloh, Ryan and Teague in 2013~\cite{DBLP:conf/spw/BenalohRT13}, several hybrid schemes have been proposed in the literature. Of course, building such a hybrid system requires trade-offs, and balancing different requirements may lead to several possible equilibria. 

In this paper, we are going to take a closer look at a recent proposal presented by a team of Belgian researchers with the aim of being implemented for postal voting in Belgium~\cite{BelgianVoting,Abeels2021,Devillez22}. The advantage of this proposal over the previous purely academic papers is that it comes with technical implementation details much better laid out. Belgium also has a national electronic identity system which opens up new opportunities in term of voter authentication and eligibility verification.

The paper is organized as follows. We will review verifiability properties and some of the previous verifiable postal voting schemes in Section~\ref{sec:SotA}. Section~\ref{sec:analysis} summarizes the main properties of the proposed Belgian system and gives an analysis of some of its shortcomings. Next, Section~\ref{sec:blindsig} proposes possible improvements and studies their pros and cons. Section~\ref{sec:discussion} provides further discussion, and finally Section~\ref{sec:conclusions} draws some conclusions.

\section{State of the art -- verifiability properties and verifiable postal voting}
\label{sec:SotA}

Integrity of the tally is one of the central security requirements for voting. In order to achieve this, voting systems provide various \emph{verifiability} mechanisms. They may vary in terms of who can perform the verification, and validity of which claims is ensured as a result. There is significant body of research concerning different definitions and the relationships between the corresponding notions; see~\cite{DBLP:conf/voteid/Kusters017,DBLP:conf/sp/CortierGKMT16} for good overviews on the topic.

A typical goal for voting systems is \emph{end-to-end} (E2E) verifiability that can be viewed as a combination of three more specific properties; see e.g. Crimmins \emph{et al.}~\cite{https://doi.org/10.48550/arxiv.2111.08662}:
\begin{itemize}
    \item \emph{Cast as Intended} requirement means that the votes reach the system (i.e. a physical or digital ballot box) the way the voter intended to.
    \item \emph{Counted as Cast} property aims at ensuring that the votes found in the ballot box are processed and tallied correctly. 
    \item \emph{Eligibility Verifiability} is meant to guarantee that only eligible voters are allowed to vote, i.e. no ballot box stuffing, double voting, etc. has occurred. 
\end{itemize}

It is also important to consider who is capable of running the verification procedures. 

We say that a property is \emph{individually verifiable} is the voter herself can get assurance that the desired property holds. At the very least, Cast as Intended requirement should be individually verifiable as otherwise the voter would get no integrity guarantees at all. The Counted as Cast would be very nice to verify individually as well, but proper care must be taken when providing such a capability in order for not to enable vote selling or other coercive practices. Eligibility of all the voters is, however, something that is very hard for every individual to verify.

Hence it is also important to have system-side auditors in a voting system. We say that a property is \emph{universally verifiable} if it can be ensured by a designated auditor. A typical example here is verification of the Counted as Cast requirement, be it physical recount of the ballots in a polling station or checking zero-knowledge decryption proofs of an electronic voting system.

Crimmins \emph{et al.} also identify a third type of  verifiability~\cite{https://doi.org/10.48550/arxiv.2111.08662}. They say that a property is \emph{verifiable collectively} if a group of voters can get sufficient assurance that the property holds. It can be viewed as an extension to individual verifiability. If, for example, the system can cheat the verification attempt by a single voter with probability $\frac{1}{2}$, a probability that it successfully cheats $k$ independent voters is $2^{-k}$.

Means to achieve these properties depend on the features of the system, readiness of the electorate, limitations in regards to other requirements, etc.~\cite{DBLP:conf/acns/PankovaW22}

For example, in order to achieve universal verifiability, voting schemes foresee various checks such as re-counting or risk-limiting audits~\cite{DBLP:journals/ieeesp/LindemanS12}. However, these techniques are typically only available for designated auditors and not regular voters. To address this issue, several ideas to extend verification options for individual voters have been put forward in various settings, most notably in-person polling site vote casting (see~\cite{ali2016overview} for a good overview). 

In this paper, however we concentrate on pure postal voting. We will limit the treatment to the systems that use only regular computer hardware for at-home paper ballot printing. Thus, the systems relying on special features like scratch surfaces in style of Remotegrity~\cite{DBLP:conf/acns/ZagorskiCCCEV13} or SAFE Vote~\cite{https://doi.org/10.48550/arxiv.2111.08662} remain out of scope of the current paper.

Of course, also in the case of home-printed ballots we would like to achieve individual, universal and end-to-end verifiability, respecting vote secrecy as much as possible in the process. There are a few solutions allowing to make different trade-offs between these properties described in the existing literature.

In 2013, Benaloh, Ryan and Teague proposed a system (which we will subsequently call BRT) with a verification process reminding randomized partial checking~\cite{DBLP:conf/spw/BenalohRT13}. Each ballot contains a plaintext vote, an encrypted and signed representation of the vote, and cryptographic material allowing to make a link between the two. The key idea is to publish the connection between the plaintext vote to the linking material for half of the votes, and the connection between the linking material and the encrypted and signed votes for the other half. If the random sample generation process is fair and verified, the published links allow to certify correctness of the tally. 

In 2021, Benaloh proposed STROBE, a system providing the voter with two blank ballots having independent sets of verification codes~\cite{DBLP:conf/evoteid/Benaloh21}. The voter picks one of the ballots, fills it out and submits. Verification codes on this ballot allow the voter to verify that her vote has been received as intended. For the other ballot, however, decryptions and randomness used to generate all encryptions on it will be published, allowing the voter to check that ballot encoding has been performed properly. Thus the ballot creation application has $50\%$ chance to avoid being caught cheating on a single voter. However, the system provides collective verifiability in the sense of Crimmins \emph{et al.}~\cite{https://doi.org/10.48550/arxiv.2111.08662}: if $n$ voters perform the verification independently, the cheating ballot creator is caught with probability $1-2^{-n}$. 

It was observed by Crimmins \emph{et al.} that STROBE poses a number of practical implementation problems such as the double ballot system being error-prone from both the voter's and the system's perspective~\cite{https://doi.org/10.48550/arxiv.2111.08662}. As improvements, they proposed two alternatives. 

In case of RemoteVote, two sets of short verification codes are printed on the same ballot, and after delivery, one of these columns is selected verifiably randomly by the system. Then the voter can use a ``trusted third-party system to generate a partial image for their ballot [that] would display one column of expected shortcodes next to the appropriate candidates''. The paper~\cite{https://doi.org/10.48550/arxiv.2111.08662} is low on details how exactly this would happen, but we argue that the result is not necessarily less error-prone than the original STROBE system.

Crimmins \emph{et al.} also propose an alternative code-named Scratch Auditing for Fair Elections (SAFE) Vote. The idea is to hide a cryptographic key under a scratch surface which can be removed to perform the ballot correctness audit. The voter can then request another ballot (many times, if she so wishes), or choose to submit a scratched one. In the case of an unscratched submitted ballot, auditing information would be displayed on the bulletin board. Again, such a system would not necessarily be less error-prone than STROBE, and the need for scratch surfaces would prevent digital ballot delivery.

Perhaps the cryptographically most involved scheme to date has been proposed by McMurtry \emph{et al.}~\cite{https://doi.org/10.48550/arxiv.2111.04210}. In order to provide vote and tally integrity verification, the voter's device computes and publishes a Carter-Wegman hash of the vote, together with the commitments to the secret values used during the hash computation. The hard part in their protocol is eligibility verification during the tally auditing. It requires identity information to be stored in special double envelopes, and handled by the tallying authority so that no-one would physically look at the identifying data while opening the envelopes.

Most of the above schemes (except for SAFE-Vote) foresee an option of digital ballot delivery, but the approaches to ballot generation and marking vary. 

In case of BRT and the system by McMurtry \emph{et al.}, ballot generation and filling happen on the voter's device. 
STROBE originally proposes blank ballot generation on the server side, but as an extension, the option of generating the blank ballot on the user device is also discussed~\cite{DBLP:conf/evoteid/Benaloh21}. RemoteVote delivers an initial paper ballot, but the ballot that actually gets submitted is generated by the voter's device in collaboration with the server.

Remarkably, none of the above systems explicitly deals with eligibility verification, even though it is one of the three core components of E2E verifiability.

BRT essentially takes a PKI approach, assuming that ``\emph{there is a public list linking a public key to each eligible voter.}''~\cite{DBLP:conf/spw/BenalohRT13}. The corresponding private key is used to create an encrypted and signed digital envelope ``\emph{which identifies (in non-human-readable form) whose key it is signed with.}'', When the votes are processed, ``\emph{for each envelope, the signature [- - -] is verified without revealing to observers whose signature it is.}'' We argue that this is a process easy to draft on paper, but difficult to establish a practical implementation for. The paper leaves it unclear what exactly are the eligibility observers supposed to be convinced of, and how to balance eligibility verifiability with coercion-resistance. Note that the observer can also be malicious and may be willing to use technical means to interpret non-human-readable values.

McMurtry \emph{et al.} state that ``\emph{For eligibility verifiability, we need to assume that the
public has some way of assessing whether a VoterID corresponds to an eligible voter.}''~\cite{https://doi.org/10.48550/arxiv.2111.04210}, but they do not discuss the methods of ensuring that this assumption holds. Note that a simple solution of publishing the mapping between the VoterID-s and a (public) list of eligible voters does not work well. This would leak the list of voters who actually cast theirs vote, enabling efficient verification mechanism for the coercive attack of forced abstention. Venice Commission clearly labels this as an unfavorable practice in Article I.4.54 of the Code of Good Practice in Electoral Matters~\cite{Code-of-Good-Practice} stating that ``\emph{Moreover, since abstention may indicate a political choice, lists of persons voting should not be published}''.

Crimmins \emph{et al.} claim in~\cite{https://doi.org/10.48550/arxiv.2111.08662} that STROBE, RemoteVote and SAFE Vote all provide the eligibility verification property, but do not specify how exactly. The only explanation given is a reference to 'existing procedural controls' in a footnote, possibly hinting at the standard methods used in postal voting like double envelopes. 

Note, however, that double envelope system is a legacy adopted not because of its excellent properties, but because historically there has not been a better alternative. For instance it does not really protect vote secrecy against a malicious actor while the vote is in transit; thus we question the ballot secrecy claims made in Table 1 of~\cite{https://doi.org/10.48550/arxiv.2111.08662}. This is a good example that one can not leave any part of the system unspecified while proposing a new voting scheme as implementation details of one component may harm the desired properties of others.

\section{The proposed Belgian remote voting system}
\label{sec:analysis}

In the current paper we will concentrate on a very interesting initiative recently taken by the Belgian election authorities. A study published in 2021 analyzes the options of introducing remote electronic voting in Belgium~\cite{BelgianVoting}. For the time being, it is advised not to start the development efforts for vote casting via Internet, but a number of proposals are made to improve security properties of the currently used postal voting. 

\begin{figure*}[ht]
    \begin{center}
        \includegraphics[width=12cm]{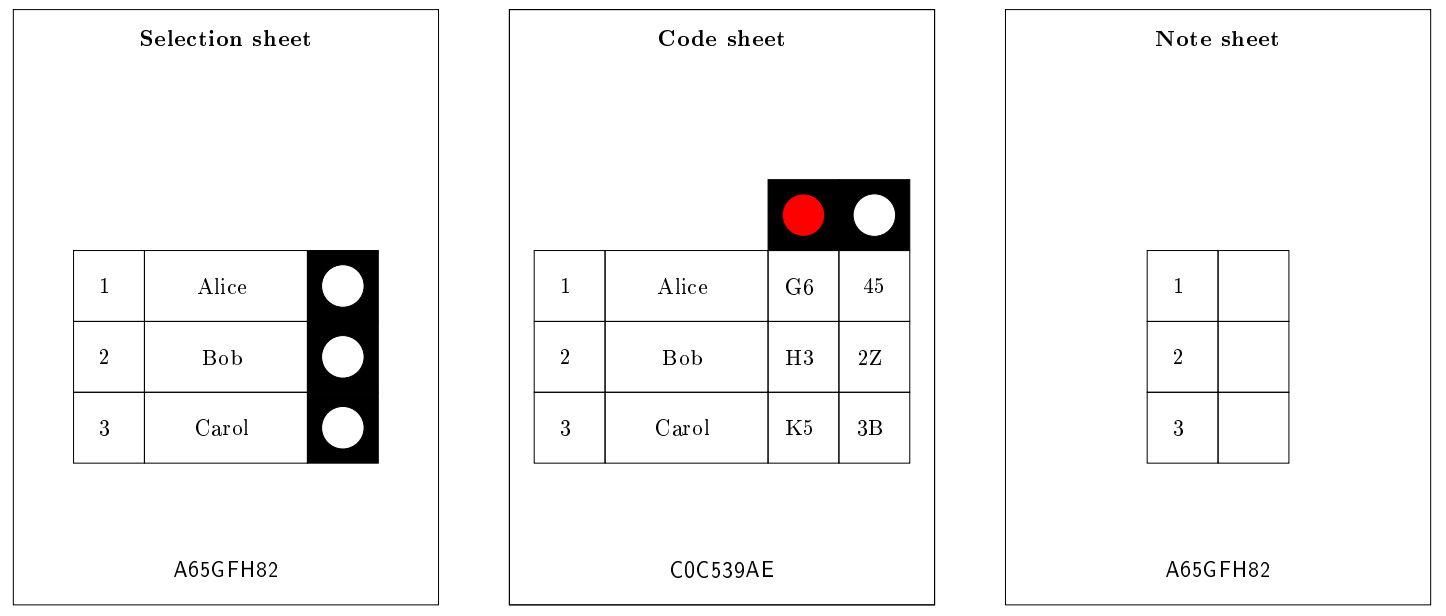}
        \caption{Ballot, code and note sheet for the proposed Belgian postal voting system~\cite{BelgianVoting,Abeels2021,Devillez22}}
        \label{fig:belgian}
    \end{center}
\end{figure*}

The system proposed for Belgium also relies on the verification codes that have to be recorded by the voter in order to perform the verification later~\cite{BelgianVoting,Abeels2021,Devillez22}. More precisely, the voter is provided with three sheets (see Figure~\ref{fig:belgian}). The selection sheet lists all the candidates together with the preference marking spots. The code sheet presents short codes for both of the options of voting for or against a particular candidate. Finally, the note sheet is meant for the voter to write down the codes corresponding to her selections in order to later check against the codes published on the bulletin board. As the code sheet provides a receipt of voting, it is meant to be destroyed after the vote has been cast.

What makes the case of the proposed Belgian system especially interesting is the existence of a well-established national electronic identity (eID) infrastructure. It means that  voter authentication can be performed much more reliably, potentially also improving eligibility verification in the case of postal voting.

In the beginning of the process, the voter logs onto the voting server by using her eID. The server checks eligibility and prepares the voting sheets specifically for this voter. Each sheet will carry a random 128-bit voter-specific code $k$, enabling the tallying authority to authenticate the vote even without relying on an outer, signed envelope. This allows, in principle, to drop the outer envelope altogether, thus potentially increasing postal vote secrecy during the vote transmission.

In Belgium, the selection sheets can be pretty large as they need to accommodate all the candidates. However, the voter can only vote for the candidates of one party. Thus, as a compromise, in case of electronically prepared ballots, it is proposed that the candidates of one party are displayed on one A4 paper, and the voter would need to mail in only the sheet corresponding to the party of her choice.

It has been left unspecified in the system description~\cite{BelgianVoting} whether the voter can only print out the sheet she needs for her party of choice, or whether she should print out all the generated sheets. The subtle issue here is keeping the vote secret from the voter's computer. If the voter would choose to print the candidate list of only one party, a malicious device would learn her party preference.

There are a few ways to look at the issue. On one hand, voter's device definitely is one of the easiest-to-attack components in the whole system, especially when it comes to vote secrecy. Even if the system provides vote integrity verification mechanisms, it is hard to give strong guarantees that the vote has not been leaked from the used digital device. The best known mechanism to achieve such guarantees would be code voting, but it comes with usability trade-offs~\cite{DBLP:conf/chi/MarkySLM19} and we do not consider such systems in this paper.

A usual approach to this problem is not to target vote secrecy at all in remote settings, and aim at a weaker property of coercion resistance instead (see e.g.~\cite{DBLP:conf/voteid/KripsW19} for an overview of different proposed approaches to achieve it). Since the Belgian system has been presented as the first step of transition towards Internet voting, there will be a moment in the future when the voters will use their computing devices to prepare and cast votes. Thus we argue that leaking one's vote to the computer is a practical trade-off that will need to be accepted at some point anyway.

We note that in the Belgian system as it is described in~\cite{BelgianVoting}, the voter has to trust her computer also in regards to vote integrity. It is foreseen that the voter can contact the ballot preparation server and check that the code $k$ is a valid one, but it is not guaranteed to be unique. If an attacker is able to compromise several voter devices, he can make these devices to use the same (valid!) $k$ for all the ballots issued through them. This problem would only be noticed in the tallying phase and the system description~\cite{BelgianVoting} does not specify what to do in this case. However, there are little alternatives to invalidating all the votes sharing the same $k$, as the tallying authority can not distinguish the $k$-sharing-attack from a ballot box stuffing attempt. This efficiently results in disenfranchising all the voters who cast these votes.

We may try to detect multiple verification attempts made to the same code~$k$, but it is unclear what to do in case of successful detection. The voter may legitimately want to verify the code several times from different devices as she does not necessarily trust a single device. Also, most of the voters would probably not bother verifying the code at all, and thus such a detection mechanism would likely be inefficient.

We also note that checking the value of $k$ for validity may pose a usability issue. The system description~\cite{BelgianVoting} discusses embedding $k$ on the ballot sheets both in an OCR font and in the form of a QR code, recommending the former to support human readability. In both cases, the voter would need a device capable of scanning the representation of $k$, which in the current practice means having a smartphone, a tablet computer or alike. In any case the success of scanning depends on the user skills, quality of the camera, lighting conditions, etc.

\section{Eligibility assurance with blind signatures}
\label{sec:blindsig}

The root of the problem enabling reuse of the values for $k$ is that these values depend neither on the voter, nor the vote. Of course we do not want to print the voter's digital signature on the ballot instead as this would undermine vote secrecy. Luckily, there exists a good alternative available in the form of blind signatures.

Blind signatures were first introduced by Chaum in 1982 in the context of implementing untraceable payment systems~\cite{DBLP:conf/crypto/Chaum82}. In 1992, Fujioka \emph{et al.} proposed using this primitive to achieve vote-secrecy-preserving authentication of a ballot by blindly signing it with the authority's key~\cite{DBLP:conf/asiacrypt/FujiokaOO92}. 

The construction of Fujioka \emph{et al.} is a very generic one, with a number of improvements proposed throughout the years (see e.g.~\cite{schmid2008blind} for a good overview on the topic). Blind signatures have been used also in practical e-voting schemes; recently e.g. in Russia~\cite{DBLP:conf/eicc/VakarjukSW22}. However, their practical applicability to postal voting has been very limited. This can be explained by a diverse set of assumptions that one would need in order for make such a solution to be useful and work.

On one hand, in order to make use of blind signatures, a relatively advanced digital infrastructure is required. The voters need a reliable means for authenticating themselves to the signing authority, accompanied with a method to do something with the returned signature. On the other hand, even though a digital identification infrastructure is assumed, the society should still look to improve remote paper vote casting, rather than going for Internet voting right away.

Both of these aspects are present in Belgium, and hence considering the blind signatures for authenticity and eligibility assurance is interesting in this case.

Of course, we would need to use the voter's computer as a ballot marking device and trust it for vote secrecy. However, as discussed in Section~\ref{sec:analysis}, this is a trade-off that is probably required sooner or later anyway.


Thus, we propose setting up a generic blind signature scheme as an addition to the proposed Belgian postal voting system.
For that, we will assume the authority $A$ who maintains the list of eligible voters, possesses a public-private blind signature key pair, and publishes the corresponding public key.


After the voter has used her computer to fill in the ballot, it is first masked for blind signing. The voter then authenticates herself to $A$ who verifies her eligibility. If this verification succeeds, $A$ issues the blind signature. Next, the voter's computer removes the blinding and displays the obtained signature directly on the ballot, e.g. as a QR code. The resulting sheet can then be printed out and cast as a regular postal ballot.

Before mailing it off, this scheme allows the voter to check well-formedness of the ballot and the signature of $A$. First of all, note that the Belgian ballot can be encoded rather efficiently. There are less than $256$ parties running, so one byte is enough to encode the party choice. For each candidate of this party, one bit needs to be encoded. Depending on the length of the the party list, one may need a few dozens of bits. Adding the metadata concerning the election event, the encoding should comfortably fit into $256$ bits.

This means that we can put the padded encoding of the vote directly under the signature without hashing it. Thus, a mobile verification app can be developed that can decode the whole vote together with $A$'s signature from the QR code, check the signature and display the decoded vote content to the voter. The voter can then visually match the result to what has been printed out on the ballot in the traditional human-readable way. This ensures the voter that the vote has indeed been correctly signed by $A$ without intermediate manipulation.

Machine-readable votes also allow for a more efficient tallying process by scanning the QR codes. It is not even necessary to visually inspect all the postal ballots for correspondence to the human readable part if a proper statistical post-election audit process like RLA is implemented. Note that a statistical post-election audit as part of the tallying procedures implicitly also protects the voters who did not bother downloading and using the mobile verification app.

As the ballot is signed with the authority's signature to prove eligibility, there is no need for the outer, voter-identifying envelope, and the ballot can be mailed anonymously. This removes one of the major privacy problems of postal voting that anyone can study the envelopes in transit and reveal how the postal voters voted. At the same time, blind signatures printed on the ballots ensure eligibility of the voters and also protect ballot integrity.

On the other hand, extra measures are then needed at the polling station on the election day. If a person who has cast a postal vote comes to the polling station and wants to cast a vote, a respective mechanism is needed to avoid double voting. 

In case of a standard double envelope postal voting system (see e.g.~\cite{DBLP:conf/voteid/KillerS19}), envelopes can be kept sealed until the regular polling station votes are also cast. Double envelopes belonging to the voters who submitted in-person votes can then be discarded without opening. 

In case of anonymously sent postal votes this approach would not work. Instead, the voter needs to be stopped at the polling station before she gets a chance to submit a vote. For that, polling stations workers need access to $A$'s database of voters who have requested signing their postal votes. This is technically non-trivial, but doable. A similar system has been in use in Estonia since 2021 Parliamentary elections to enable cancelling electronic votes with paper ones in a polling station~\cite{EHIN2022101718}.

Note that the problem of double voting is also present in the proposed Belgian system as described in~\cite{BelgianVoting}, and even on a bit more serious level. In principle, the ballot preparation server can keep a list of the voters who have requested a ballot, but it can not tell if the ballot has actually been completed. If requesting a blank ballot would be registered as the voter having used her voting rights, this may end in disenfranchising the voter e.g. in the case she fails submitting her postal vote and attempts voting in a polling station. 

In case of our proposal, the voter only requests the authority's signature \emph{after} having filled the ballot in. Of course, we still do not know whether the signed ballot was actually mailed or not. However, we argue that there is a potential legal difference between just requesting a blank ballot and asking for the authority's confirmation once it has been filled. In the latter case it is easier to call the act of voting completed and rule against the voter in case of a possible dispute between disenfranchisement \emph{vs.} double voting.

Using the voter's computer as a ballot marking device also allows for a more efficient printing procedure. There is no reason to print all the sheets corresponding to the parties the voter did not want to vote for. Of course, this is mainly a result of our trade-off with secrecy of the vote from the voter's computer. On the other hand, it also gives a significant environmental effect as the number of otherwise unused sheets of paper would be multiplied by the number of postal voters.

Note also that we do not need to print the code sheet at all. Instead, we can directly generate and print the filled note sheet. This is good both from the usability and security points of view. Usability benefits are clear as the voter is not required to copy any random codes by hand. Security benefit comes from the observation that the code sheet is actually a receipt that the voter can use intentionally or under coercion to prove how she voted. 

The original system description~\cite{BelgianVoting} requires the voter to destroy this sheet, but we argue that relying on such a measure to achieve privacy properties is not a good security design principle. Users can in general only be expected to give a minimal amount of effort to achieve the functional goals, i.e. casting one's vote in our case. If the code sheet remains lying around, it can cause unexpected privacy problems which are better avoided if possible.

\section{Discussion}
\label{sec:discussion}

Verifiability properties of standard double envelope postal voting are rather weak. There is typically no Cast as Intended verification, and instead of Counted as Cast there is a weaker property of Counted as Collected~\cite{DBLP:conf/voteid/BernhardBHRRSTV17}. We argue that if such a system would be proposed today, it would not be accepted as not satisfying elementary requirements, especially as postal voting protocols offering better properties are available now~\cite{DBLP:conf/spw/BenalohRT13,DBLP:conf/acns/ZagorskiCCCEV13,DBLP:conf/evoteid/Benaloh21,https://doi.org/10.48550/arxiv.2111.04210,https://doi.org/10.48550/arxiv.2111.08662,BelgianVoting}.

However, eligibility verification remains a challenge for all these proposals, and this problem is inherently related to the available infrastructure. When we want to enable e.g. Cast as Intended verification, we need to enhance the capabilities of the verifier, i.e. the voter. When postal voting was introduced for the soldiers fighting in the U.S. Civil War, there was no way of getting convenient and fast feedback about the fate of the vote~\cite{CivilWar}. But nowadays we have omnipresent Internet access, enabling such feedback.

A similar situation also occurs for eligibility checking. However, now the primary verification agent is the election organizer who needs to decide whether the vote came from a legitimate voter, and whether it is a double vote or not. The ballot can carry some sort of an identifier (like a social security number), or the outer envelope may carry a signature, but neither of them can be considered a strong form of identification in the third decade of the 21st century.

In order to provide better eligibility verification properties, a respective infrastructure is required. With electronic identity mechanisms being introduced in many countries, this infrastructure is becoming readily available. It is only natural to use it to secure remote voting, both in electronic and mail-in settings.

The most straightforward way of integrating an eID into a remote voting scheme would be signing the vote. In case of electronic voting it is easy to encrypt the vote in order to protect its confidentiality. For postal voting, however, there is an implicit expectation that the paper representation of the vote should be human-readable. This makes direct signing with voter's eID impossible.

On the other hand, blind signature on an anonymous paper vote is still very much an option. Of course, a corrupt signing authority may attempt to sign the votes for ineligible voters. As a solution, we can require the blind signing requests to be signed by the voters. If in the end of the voting period the number of authority-signed votes in the digital ballot box exceeds the number of voter-signed requests then we know that the authority has cheated. As an alternative, signing authority can be implemented in a distributed manner in order to avoid relying on just one trusted party.

Besides the Belgian voting system, blind-signatures-based eligibility assurance can also help to improve auditability and vote secrecy properties of other schemes. In case of BRT, for example, a malicious auditor would no more be able to breach vote secrecy in the tallying phase. For the proposals that have largely ignored laying out the details of eligibility verification (including STROBE~\cite{DBLP:conf/evoteid/Benaloh21}, RemoteVote~\cite{https://doi.org/10.48550/arxiv.2111.08662}, and the scheme by McMurtry \emph{et al.}~\cite{https://doi.org/10.48550/arxiv.2111.04210}), blind signatures provide one possible implementation of the protocol for the auditors to check voter eligibility in a privacy-preserving manner. Besides better auditability, all of the considered schemes would be protected against vote privacy breaches while the votes are in transit, as opposed to the potential alternative of the double envelope system.

We also note that while the idea of using blind signatures in a remote voting setting is not novel, their application to paper-based voting systems is to the best of our knowledge.

\section{Conclusions}
\label{sec:conclusions}

Cryptographically-enhanced postal voting is a recent and exciting research area. It has a potential to provide a remote voting solution with better authentication and integrity properties compared to regular postal voting. At the same time, it can also avoid some of the problems with remote electronic vote casting as the main vote carrying medium would still be paper.

In this article we reviewed several recent schemes, concentrating on the system proposed by Belgian researchers as an intermediate step towards Internet voting. It adds E2E verification capabilities to the postal votes and can even be used to send the filled ballots in anonymously.

The proposal is very detailed in implementation details compared to previous purely academic papers. It is also very realistic in terms of the trade-offs required between usability, verifiability and privacy properties of the system. 

However, we were still able to point out several problems in this paper. The biggest issue is the need to trust the voter's computer not to disenfranchise the voter by maliciously re-using the random authentication token $k$. 

In order to mitigate this problem, we proposed implementing a generic blind signature scheme instead of using the random token. It turns out that such a solution also has other benefits; for example it enables easier tallying and vote correctness verification by the voter. 

The downside of our proposal is the need to use the voter's PC as a ballot preparation device, hence trusting the device not to breach vote secrecy. However, we argue that this is a reasonable trade-off that will need to be accepted at some point anyway. At the same time we reduce the need for paper print-outs. This improves both the environmental footprint and coercion-resistance properties of the scheme.

The Belgian postal voting scheme is still in the early stages of research, and we hope that this paper has made a small contribution towards its future success.

\section*{Acknowledgments}

The paper has been supported by the Estonian Research Council under the grant number PRG920.

\bibliographystyle{splncs04}
\bibliography{BlindSigPostal}

\end{document}